\documentclass[3p,times,twocolumn,11pt]{elsarticle}

\usepackage{amssymb}
\usepackage{color}
\definecolor{Blue}{rgb}{0.3,0.3,0.9}

\journal{the International Journal of Non-Linear Mechanics}

\newcommand{\beq}{ \begin{equation} }
\newcommand{\eeq}{ \end{equation} }
\newcommand{\beqs}{ \begin{eqnarray} }
\newcommand{\eeqs}{ \end{eqnarray} }
\newcommand{\f}[2]{\frac{#1}{#2}}

\newcommand{\Dp}[2]{\ensuremath{\f{\partial#1}{\partial#2}}}

\begin{document}

\begin{frontmatter}



\title{Dynamics of digging in wet soil}


\author[label1]{Sunghwan Jung}
\author[label2]{Amos G. Winter, V} 
\author[label2]{A.~E.~Hosoi}
\address[label1]{Department of Engineering Science and Mechanics, \\
Virginia Polytechnic Institute and State University, 
Blacksburg, VA 24061,  USA }
\address[label2]{Department of Mechanical Engineering, Hatsopoulos Microfluids Laboratory \\
Massachusetts Institute of Technology, Cambridge, MA 02139,  USA}

\begin{abstract}
Numerous animals live in, and locomote through, subsea soils. To move in a medium dominated by frictional interactions, many of these animals have adopted unique burrowing strategies. This paper presents a burrowing model inspired by the Atlantic razor clam ({\it Ensis directus}), which uses deformations of its body to cyclically loosen and re-pack the surrounding soil in order to locally manipulate burrowing drag. The model reveals how  an anisotropic body -- composed of a cylinder and sphere varying sinusoidally in size and relative displacement -- achieves unidirectional motion through a medium with variable frictional properties. This net displacement is attained even though the body kinematics are reciprocal and  inertia of both the model organism and the surrounding medium are negligible.  Our results indicate that body aspect ratio has a strong effect on burrowing velocity and efficiency, with a well-defined maximum for given kinematics and soil material properties.  
\end{abstract}


\end{frontmatter}


\section{Introduction}

There are many examples of animals that live in particulate substrates which have evolved unique locomotion schemes \cite{Trueman75}. Two common strategies observed in biological systems are an undulatory, snake-like motion \cite{MDLG2009, Wallace1968,  Jung10, Kelly05} and a ``two-anchor" system \cite{Fager64, HD77,  SNC02, Stanley69, Trueman66, Trueman67, TBD66}.  An example of the former is the sandfish lizard which wiggles its body from side to side in order to effectively swim through sand \cite{MDLG2009}. Similarly, smaller organisms like {\it C. elegans} have been observed to move quite efficiently via an undulatory motion through granular media \cite{Wallace1968,Jung10}. In contrast, soft-bodied organisms that live in particulate substrates saturated with a pore liquid generally use a two-anchor system to burrow. 
In this strategy, one section of the animal expands to form a terminal anchor, while another section of the animal contracts to reduce drag. Once the contracted section is conveyed forward in the burrow, it is expanded to form the next terminal anchor and the previous terminal anchor is contracted and shifted forward. 

The burrowing model presented in this paper is inspired by the two-anchor locomotion scheme and body geometry of the Atlantic razor clam ({\it Ensis directus}). {\it Ensis} is comprised of a long, slender set of valves (i.e.~the two halves of the shell) which are hinged on an axis oriented longitudinally to the animal, and a dexterous soft foot which resides at the base of the valves. The burrowing cycle of {\it Ensis} is depicted in Figure \ref{Fig_Clam_motion} (a). The animal starts with its foot fully extended below the valves (A). Next, it uses a series of four shell motions to make downward progress: (B) the foot extends to uplift the valves while the valve halves contract to force blood into the foot, inflating it to serve as a terminal anchor; (C) the foot muscles contract to pull the valves downwards; and (D) the valves expand in order to form a terminal anchor and begin the cycle again. 

The uplift and contraction motion of the valves draw water towards the animal's body, unpacking and locally fluidizing the surrounding substrate \cite{winter2010thesis}. 
{\color{blue} The initiation of valve contraction causes local soil failure around the animal and the uplift velocity is on the order of the pore fluid velocity required to induce a fluidized bed below the animal.\footnote{A full description of {\it Ensis} burrowing mechanics is beyond the scope of this paper, but can be found in \cite{winter2010thesis}.} 
Although the animal is  too weak to pull its shell through static soil (which exerts a resistance that linearly increases with depth \cite{TM96}) to typical burrow depths, fluidization dramatically reduces drag, resulting in resistance forces that are depth independent \cite{winter2010thesis}. }
The aim of this paper is to analyze the kinematic motion of the shell and demonstrate that reciprocal body deformations  can produce unidirectional motion in a substrate of varying frictional properties.

\section{Model}

\begin{figure}[h]
    \centering
        \includegraphics[width=.4\textwidth]{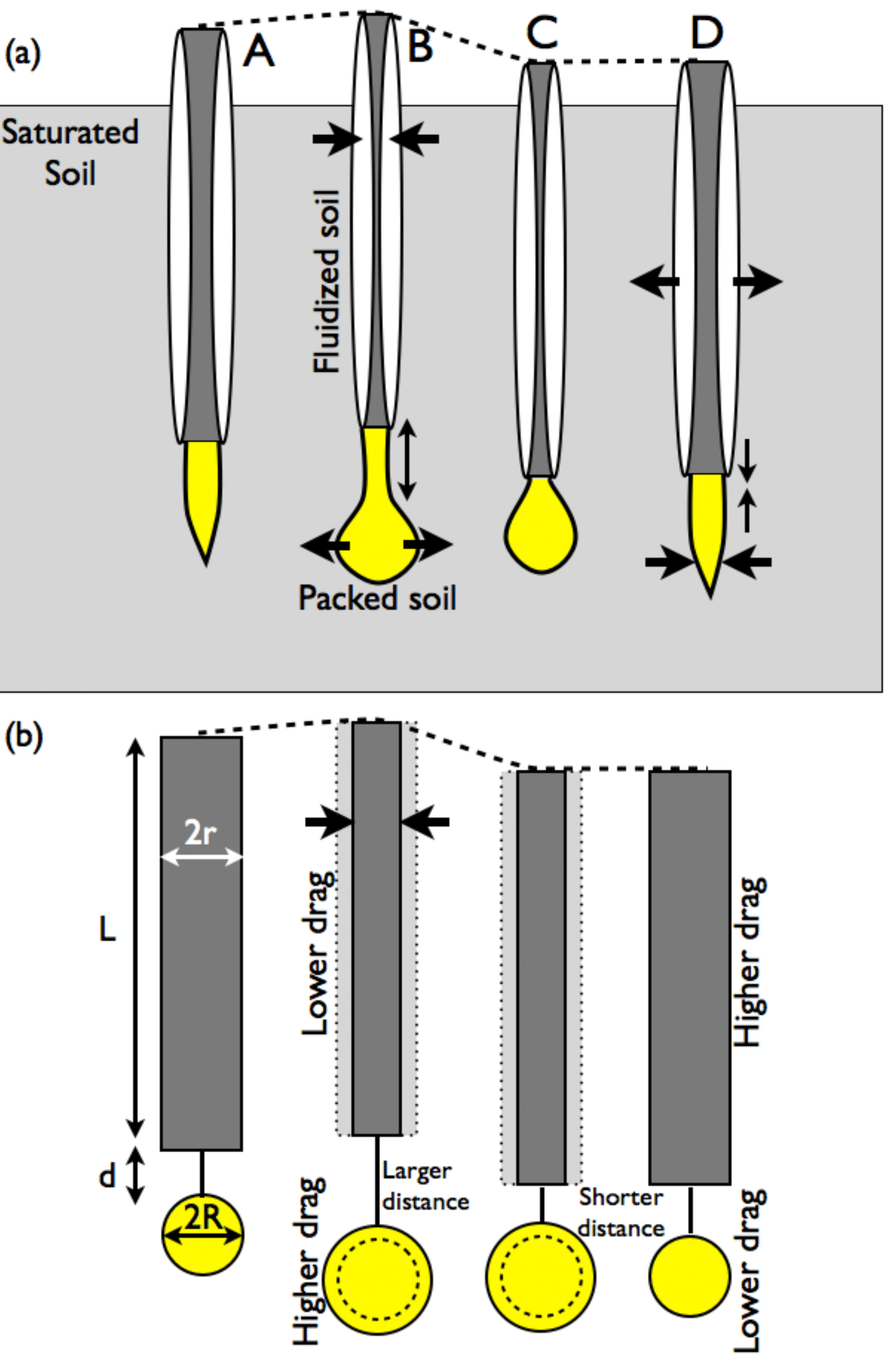}
\caption{Schematics of motion for (a) burrowing {\it Ensis} and (b) simplified model organism inspired by
{\it Ensis}. }
    \label{Fig_Clam_motion}
\end{figure}

Figure \ref{Fig_Clam_motion}(b) shows the geometry of the simplified model organism and the dynamics inspired by {\it Ensis}. The body consists of two components: a long cylinder of length $L$ and radius $r(t)$, which approximates the valves, and a sphere of radius $R(t)$ attached to the cylinder, acting as the foot.   The radius of the cylinder, the radius of the sphere and the distance between the two are known functions of time dictated by the organism.  The length of the shell, $L$, is considered constant. 

\subsection{Kinematics}

{\it Ensis}'s burrowing motions are often erratic; {\color{blue} the animal may wait anywhere from a few seconds \cite{Trueman67} to many minutes between digging cycles.} As a simplification, we consider periodic motion for both the foot and the shell. First, the radii of the cylinder and the sphere are given by
\beq
r = a_0 + a' \cos (\omega t)\, , \ \ R = a_0 - b' \cos(\omega t)
\eeq 
where $\omega$ is the frequency of motion and $a_0$ is chosen as the mean radius of both the cylinder and the sphere.

Another important factor in the kinematics of burrowing {\it Ensis} is the extension and retraction of the foot and its temporal relation to the movement of the valves. To model this motion, we impose a sinusoidally changing distance between cylinder and sphere that is out of phase with the expansions and contractions by $\pi/2$:
\beq
d ={d}_0 + d' \sin (\omega t)  \label{eq:d}.
\label{d}
\eeq

\subsection{Volume conservation}
In the live organism, the expansion of the foot is driven by fluid squeezed out of the shell. 
This fluid may be treated as incompressible at clam-like speeds and is contained in a closed loop.  Hence, by conservation of volume, as the cylinder contracts, the sphere expands and vice versa.  
For small deformations, the change in the total volume of the cylinder can be approximated as
\beq
\dot{v}_{cyl} = 2 \pi r \dot{r} L \sim - 2 \pi L \omega a_0 a' \sin(\omega t). \label{eq:vdot_cyl}
\eeq
The sphere's volume varies with opposite phase and, again for small deformations is approximated as
\beq
\dot{v}_{sph} = 4 \pi a_0^2 \dot{R} \sim 4\pi \omega a_0^2 b' \sin (\omega t). \label{eq:vdot_sph}
\eeq
Combining these two relations and applying conservation of volume shows that the amplitude of the changing sphere radius  is related to the deformations of the cylinder by $b' = (L/2a_0) a'$.

The initial mean void fraction of the surrounding medium, $\epsilon_0$,  is defined as the ratio of volume occupied by pore fluid to the total volume. 
The change in void fraction of the soil adjacent to the organism as the cylinder collapses relative to its mean value is given by
\begin{eqnarray}
 (\epsilon_0 - \epsilon) _{cyl} &=& \f{\pi L(r^2 - a_0^2)}{ {{\cal V}_{cyl}} } \nonumber \\
 &\sim&  \f{2\pi L a_0}{{ {{\cal V}_{cyl}} }} a' \cos (\omega t) \, \label{eq:eps_cyl}
\label{void}
\end{eqnarray}
where $\cal V$ is the characteristic volume of the perturbed soil, the extent of which depends on the geometry of the burrowing organism and initial soil properties \cite{WDHS10}, and $\epsilon$ is the instantaneous local void fraction. In the same manner, the void fraction change in the soil surrounding the sphere is given by
\begin{eqnarray}
 (\epsilon_0 - \epsilon) _{sph} &=& \frac{4 \pi}{3 {\cal V}_{sph} } (R^3 - a_0^3) \nonumber \\
  &\sim& - \f{4 \pi a_0^2}{{{\cal V}_{sph}}} b' \cos (\omega t) \, . \label{eq:eps_sph}
\end{eqnarray}

\subsection{Drag}
The drag force on an object moving through a saturated particulate medium depends (nonlinearly) on a number of  parameters. In general, it can be written as 
\beq
F_D = \mu(\epsilon) V^{\alpha} S \label{eq:F}
\eeq 
where $\mu$ is a resistance coefficient which depends on local void fraction $\epsilon$, 
surface roughness, shape of the body, etc., $V$ is the body's velocity, $\alpha$ is an exponent that varies with Reynolds number, and $S$ is a geometric parameter that is associated with the body's contact area with the substrate. 

For Newtonian fluids, expressions for $\mu$ in the limits of both high and low Reynolds number flows are well-known. {\color{blue} The parameter $\alpha$ characterizes the velocity dependence in the drag expression. At high speeds (or high Reynolds numbers) the drag force is strongly dependent on velocity; however, as viscous effects increase and/or the substrate exhibits increasingly solid-like behavior, this velocity dependence weakens.} 
For inviscid Newtonian flows, $\alpha = 2$, $\mu$ is related to the dimensionless drag coefficient, $C_D$, by $\mu = C_D \rho/2$ where $\rho$ is the density of the fluid, and $S$ is an effective cross-sectional area.  At low Reynolds numbers, $\alpha = 1$ and $\mu$ is proportional to the dynamic viscosity of the fluid, $\mu_f$ where the constant of proportionality depends on the geometry of the body.  For a sphere, $\mu = 6 \pi \mu_f$ and $S$ is the radius of the sphere. 

{\color{blue} It should be noted that the analysis presented in this paper is predicated on defining $\mu_f$ as an effective viscosity which correlates shear stresses to strain rate. This behavior is markedly different than critical state granular shear flow, where inter-particle frictional interactions induce shear stresses that depend on confining pressure, and are relatively independent of strain rate \cite{TM96}. Visualization of the fluidized substrate around burrowing {\it Ensis} show void fraction ranges of $0.42 < \epsilon < 0.46$ \cite{winter2010thesis}, which is unpacked beyond the point of incipient fluidization ($\epsilon \approx 0.41$) for uniform spheres \cite{wen1966}. As such, the substrate surrounding burrowing {\it Ensis} can be modeled with a fluid-like viscosity that is a function of void fraction, for which there are numerous empirically-derived expressions in the literature \cite{Einstein1906,BG1972,FA1967,KD1959}.}

Following the sequence of events depicted in Figure \ref{Fig_Clam_motion}(b), inward (collapsing) motion of the cylinder increases the local void fraction and consequently decreases the resistive drag force on the cylindrical body. In contrast, the drag on the sphere (which expands as the cylinder collapses)  increases as the cylinder collapses.  Assuming small local changes in void fraction $\Delta \epsilon = \epsilon - \epsilon_0$ about an initial $\epsilon_0$, we can write the local resistance coefficient of the cylinder, $\mu_{cyl}(\epsilon)$ as 
\beqs
\mu_{cyl}(\epsilon)
&=& \mu_{cyl}(\epsilon_0) + \left.\frac{\partial \mu_{cyl}}{\partial \epsilon}\right|_{\epsilon_0} (\epsilon-\epsilon_0) + 
 ... \nonumber \\
&\equiv& \mu_0 - \mu'_{cyl} \cos (\omega t) \label{eq:mu_cyl}
\label{resistance}
\eeqs
where $\mu'_{cyl}$ can be written as a function of the geometry of the organism and the material properties of the substrate by substituting equation (\ref{void}) into (\ref{resistance}). In general,  $\mu'_{cyl}$ is assumed to be less than $\mu_0$ otherwise the resultant negative drag on the moving body is not physical. For the sphere, $\mu_{sph}$ fluctuates out of phase with $\mu_{cyl}$ and is given by $\mu_{sph}=\mu_0 + \mu' _{sph} \cos (\omega t)$.

\subsection{Burrowing velocity}
Since we are considering an inertialess limit,  the total force on the model organism equals zero at every instant in time. 
This condition enables us to calculate a net burrowing velocity of the cylinder/sphere system.

The only forces acting on the system are drag on the cylinder and drag on the sphere which must be equal and opposite:
\beq
\mu_{cyl} |V_{cyl}|^{\alpha} S_{cyl} =  \mu_{sph} |V_{sph}|^{\alpha} S_{sph} \, .
\eeq 
Since the distance between the cylinder and the sphere, $d$, is prescribed by the digger, the two velocities must be related via a kinematic constraint: $\dot{d} = V_{cyl} - V_{sph}$, where the dot indicates a time derivative. 

\begin{figure}[t]
    \centering
        \includegraphics[width=.48\textwidth]{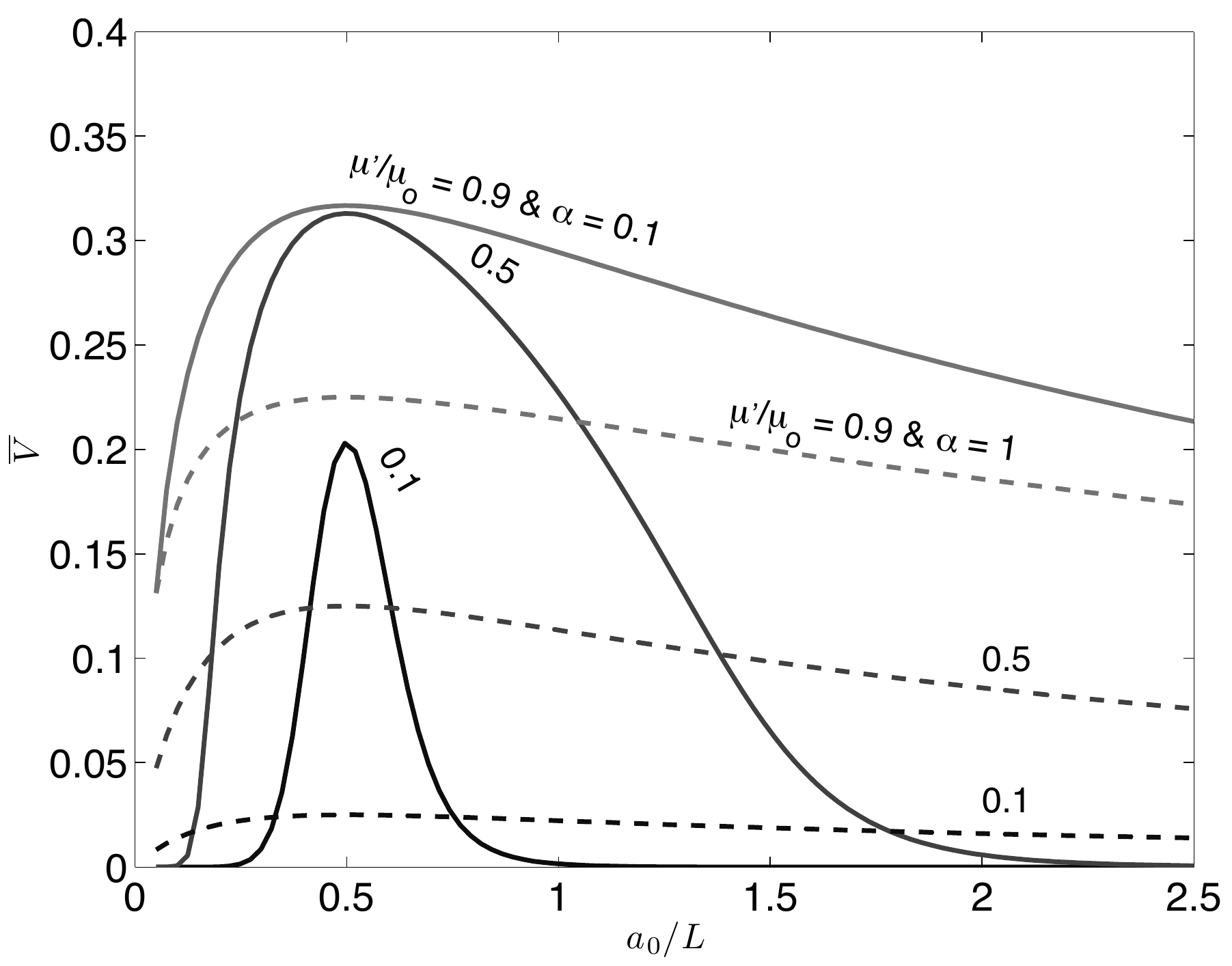}
\caption{Normalized burrowing velocity versus aspect ratio $a_0/L$ with different normalized soil parameters $\mu'/\mu_0$. Dashed lines indicate $\alpha = 1$ {\color{blue} (strong dependence on velocity)} and solid lines correspond to $\alpha =0.1$ {\color{blue} (weak dependence on velocity)}.  }
    \label{Fig_mu}
\end{figure}

\begin{figure}[t]
    \centering
        \includegraphics[width=.48\textwidth]{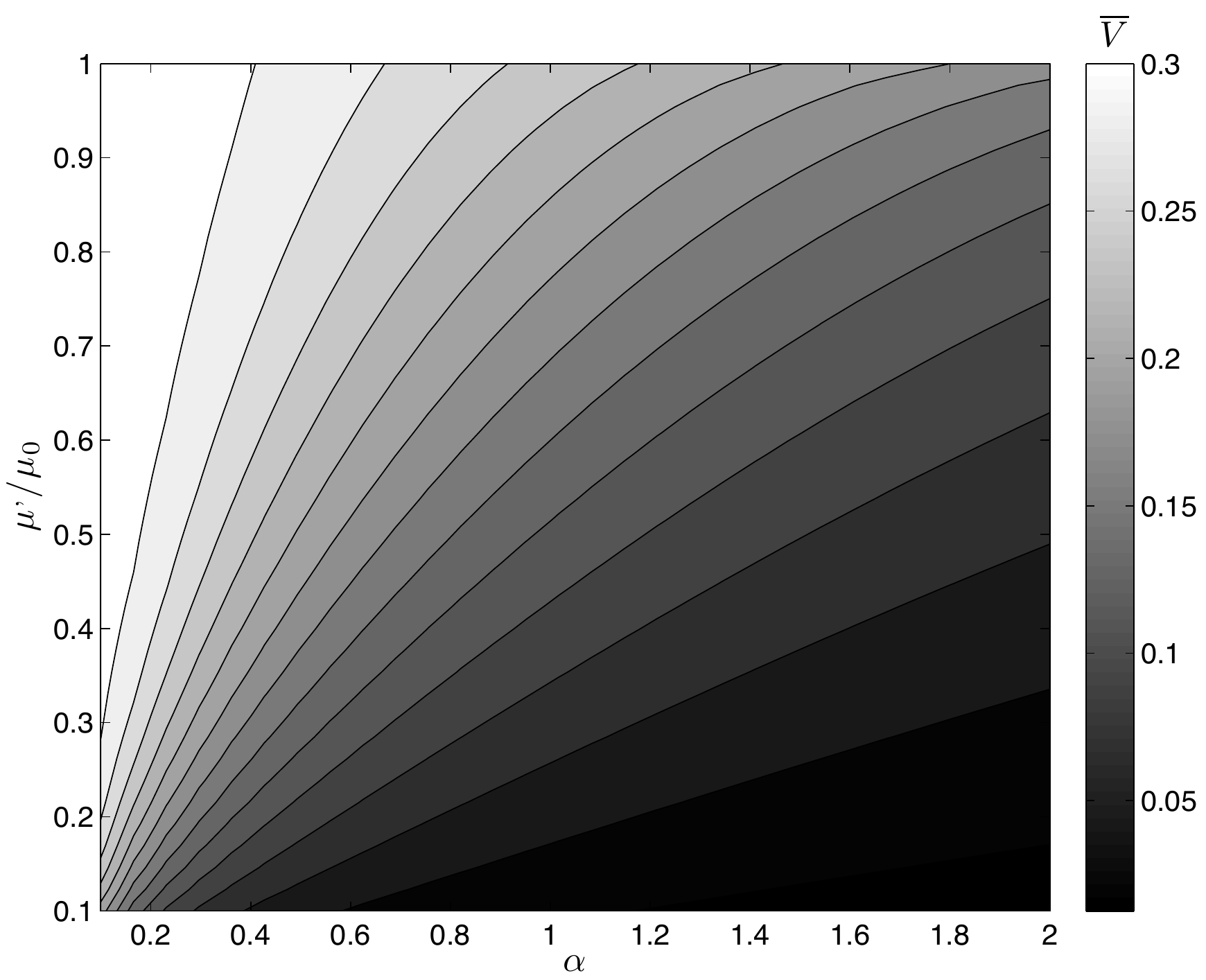}
\caption{A contour plot of maximum normalized velocity $\overline{V}_{cyl}$ {\color{blue} time averaged over one cycle} as a function $\mu'/\mu_0$ and $\alpha$ at the optimal aspect ratio, $a_0/L = 0.5$.}
    \label{Fig_contour}
\end{figure}

Solving for the velocity of the cylinder (which represents the shell of the digging clam) and substituting the kinematics defined in (\ref{d}) we find the dimensionless digging velocity of the sphere, $\hat{V}_{sph} \equiv V_{sph}/(\omega d')$:
 \beq
 \hat{V}_{sph} =  - \f{\cos (\omega t)}{1+ \gamma^{1/\alpha} } \,  
 \label{velocity}
 \eeq
where 
\beq 
\gamma = \f{ 1 + ({\mu'_{sph}}/\mu_0) \cos (\omega t)}{1 - ({\mu'_{cyl}}/\mu_0) \cos (\omega t)}  \cdot  \f{S_{sph}}{ S_{cyl} }\,.
\eeq 
Figure \ref{Fig_mu} shows the relationship between this dimensionless burrowing velocity averaged over one cycle, $\overline{V}$, and the ``shell" aspect ratio $a_0/L$ with varying normalized perturbed resistance coefficient $\mu'/\mu_0$ and varying velocity dependence $\alpha$.  
For simplicity, we have approximated $\mu'_{sph} = \mu'_{cyl} = \mu'$, and $S_{cyl}$ and $S_{sph}$ are chosen as $2\pi a_0L$ and $4 \pi a_0^2$, respectively.  Note that as an approximation, choosing $\mu'_{sph} = \mu'_{cyl}$ is a reasonable first estimate but, in reality $\mu'_{sph}$ and $\mu'_{cyl}$ are dictated by the constitutive equation relating $\mu$ and $\epsilon$ (which can be quite complicated and is often determined empirically for soils) and the geometry of the digger.  

Figure \ref{Fig_mu} indicates that, for our chosen sinusoidal kinematics,  the maximum burrowing velocity occurs at an aspect ratio of  $a_0/L = 0.5$ regardless of the material properties of the soil. 
As the normalized perturbed resistance coefficient $\mu'/\mu_0$ is increased, the maximum velocity increases and, for small $\alpha$, the  velocity profile flattens out.  
These trends indicate that, as the resistance becomes more sensitive to changes in void fraction the burrowing velocity increases. 
Figure \ref{Fig_contour} shows a contour plot of burrowing velocity as a function of a normalized perturbed resistance coefficient $\mu'/\mu_0$ and $\alpha$ at the fastest aspect ratio $a_0/L = 0.5$.
The burrowing velocity increases with decreasing dependence of body forces on local velocities and increasing perturbed resistance coefficient.

\subsection{Efficiency}

An actively burrowing animal consumes power as it deforms the surrounding medium and a certain fraction of that power is transformed into  useful unidirectional motion. 
A typical hydrodynamical efficiency can be defined as the ratio of  the power required to drag the digger through the soil at the average digging velocity (namely the useful fraction of the power) to the total power required to deform the substrate which can be expressed as 
\beq
\eta = \frac{\mu_0 \overline{V}^{\alpha+1}(S_{cyl} + S_{sph})}{\overline{\sum_{i = sph, cyl} F_D^{(i)} \cdot V_i + P_i}}
\eeq
where overline indicates a time-averaged quantity and the latter terms in the denominator, $P_{cyl}$ and $P_{sph}$, indicate the power dissipated by an expanding (or shrinking) cylinder and sphere respectively.
To evaluate the power associated with expansion, we first calculate the corresponding stresses as $\sigma_{sph} = 2 \mu_f \partial \dot{R}/ \partial r = (4/3) \mu_f \dot{v}_{sph}/v_{sph}$ and $\sigma_{cyl} = 2 \mu_f \partial \dot{r}/ \partial r = \mu_f \dot{v}_{cyl}/v_{cyl}$ where $v$ is the volume of the object and $\dot{v}$ is a dilation rate.  As before, $\mu_f$ is the dynamic viscosity of the surrounding medium which is proportional to our resistance coefficient at low Reynolds numbers. 
The power dissipated is then given by $\sigma \dot{v}$ or 
\beqs
P_{sph} &=& \frac{4 \mu_{f}}{3 v_{sph}} \dot{v}_{sph}^2 \nonumber \\
P_{cyl} &=& \frac{\mu_{f}}{ v_{cyl}} \dot{v}_{cyl}^2. 
\eeqs
Using eqs. (\ref{eq:vdot_cyl}-\ref{eq:vdot_sph}) and approximating the normalized perturbed soil parameter as the same as the normalized perturbed radius ($a'/a_0 = \mu'/\mu_0$), we can evaluate the power dissipated in expansion and contraction and compute the efficiency of digging.  Note that this approximation only considers the dissipation associated with viscous stresses generated by an expanding object and at high Reynolds numbers there are additional components associated with $P_i$. 

\begin{figure}[t]
    \centering
        \includegraphics[width=.48\textwidth]{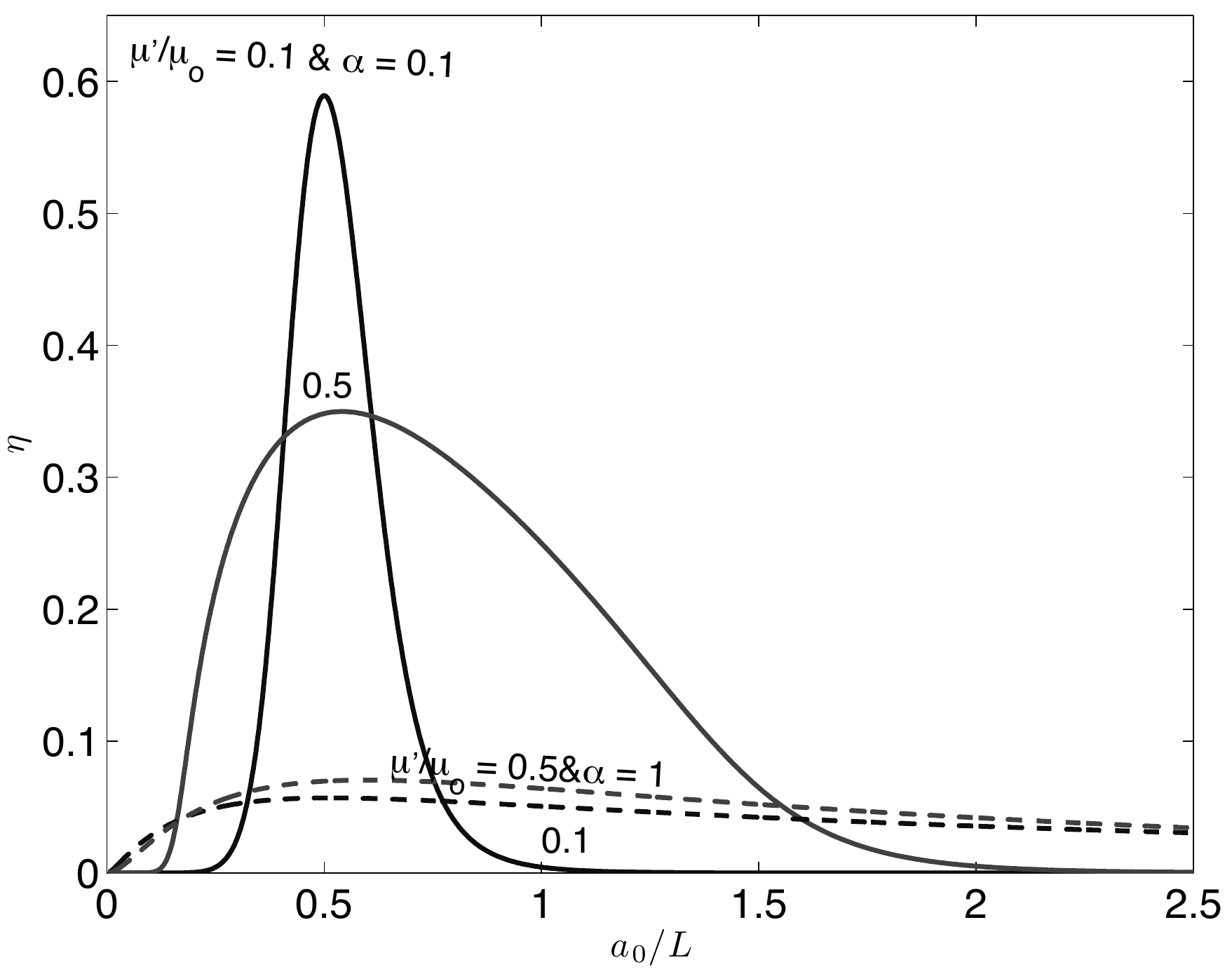}
\caption{Efficiency versus aspect ratio $a_0/L$ with varying normalized soil parameters $\mu'/\mu_0$. Solid lines indicate $\alpha = 0.1$ {\color{blue} (weak velocity dependence)} and dashed lines correspond to $\alpha = 1$ {\color{blue} (strong velocity dependence)}. }
    \label{Fig_effmu}
\end{figure}

\begin{figure}[t]
    \centering
        \includegraphics[width=.48\textwidth]{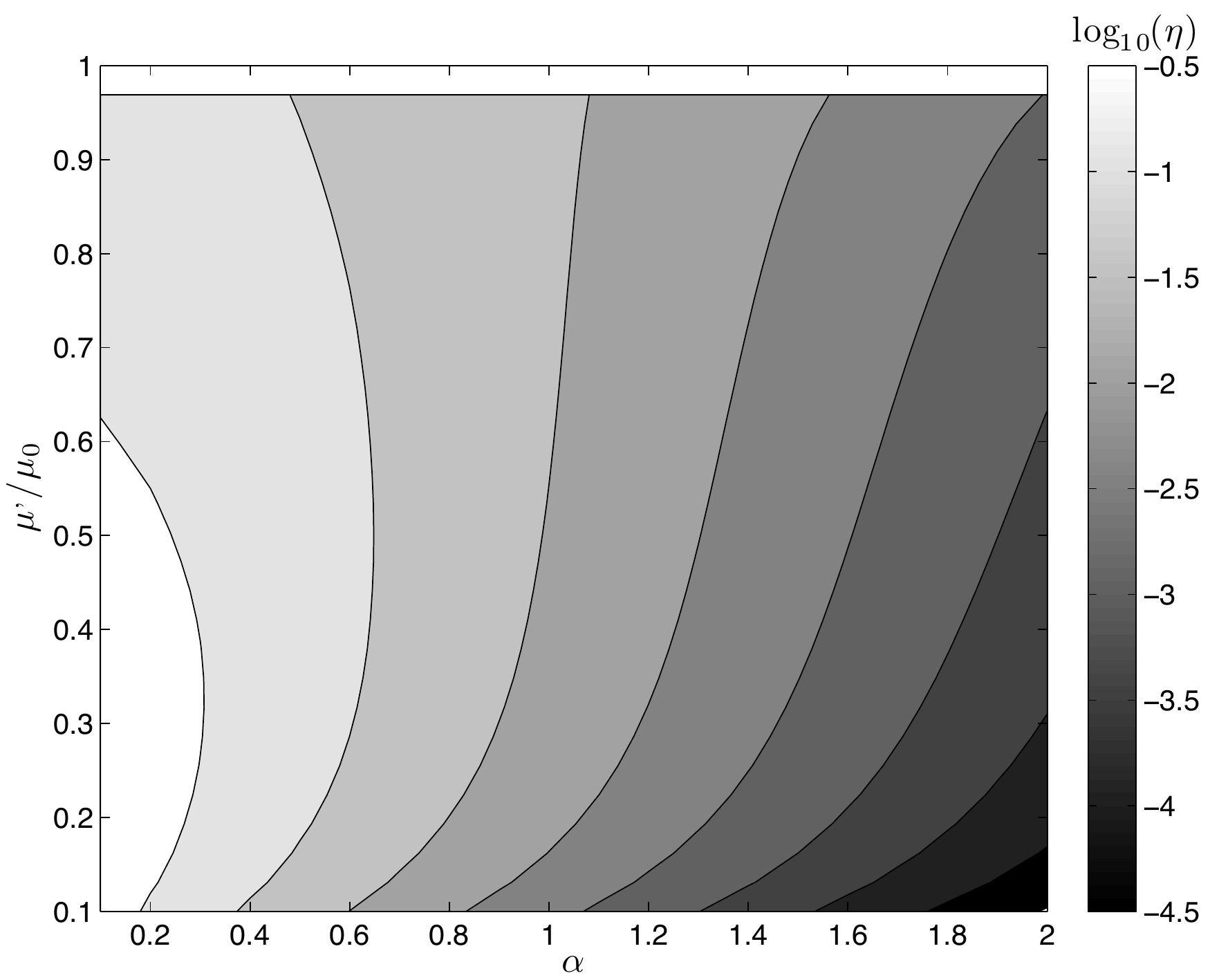}
\caption{A contour plot of efficiency $\eta$ as a function $\mu'/\mu_0$ and $\alpha$ with $a_0/L = 0.5$. }
    \label{Fig_contoureff}
\end{figure}

Figure \ref{Fig_effmu} shows the efficiency  as a function of aspect ratio $a_0/L$.  
When either the cylinder or the sphere undergoes a large variation in size, the local viscosity fluctuates significantly,
increasing the burrowing velocity. However, these large deformations also dissipate a considerable amount of energy and hence, in contrast to the velocity, the efficiency {\em decreases} as the normalized perturbed soil parameter increases.
In addition, larger normalized perturbed resistance coefficients broaden the efficiency profile as a function of $a_0/L$. 
Figure \ref{Fig_contoureff} shows a contour plot of the efficiency as a function of $\alpha$ and $\mu'/\mu_0$ at the optimal aspect ratio, $a_0/L = 0.5$. 
At small $\alpha$, the efficiency is large owing to the corresponding high burrowing velocity.
This high efficiency rapidly drops as $\alpha$ increases and varies weakly in the normalized perturbed resistance coefficient.

To compare this with other animals locomoting in a fluid environment, the hydrodynamic efficiency is roughly 0.01 for small micro-organisms in a viscous fluid, and about 0.5 for large swimming animals in an unbounded fluid.
Our results indicate that burrowing animals have relatively high efficiencies despite the large resistivity of  their surrounding environments. 

\section{Discussion} 
Given this formulation, there are a number of limiting cases that can be addressed analytically, yielding further insight into optimal geometries for burrowing. 

\subsection{Limiting case 1: Small resistance perturbations}
In the limit $\mu'/\mu_0 \ll 1$, we can estimate $\gamma^{1/\alpha}$ to first order in $\mu'/\mu_0$ as
\beq
\gamma^{1/\alpha} \sim \left( \f{S_{sph}}{ S_{cyl} } \right)^{1/\alpha}  \left( 1 + \f{2}{\alpha} \f{\mu'}{\mu_0}  \cos (\omega t) \right) \nonumber 
\eeq
were we have again approximated $\mu'_{cyl} \approx \mu'_{sph} \equiv \mu'$.
Combining this with equation (\ref{velocity}), we find the dimensionless instantaneous digging velocity can be represented as
\beqs
\hat{V}_{sph} \sim \f{1}{f} \left[ 1 - \f{2}{\alpha} \f{\mu'}{\mu_0} \f{f-1}{f} \cos (\omega t)\right] \cos (\omega t) &&\nonumber
\eeqs
where $ f(a_0/L) = 1 +  \left( S_{sph}/ S_{cyl}  \right)^{1/\alpha}$. The mean dimensionless digging velocity of the clam, namely $\hat{V}_{sph}$ time-averaged over one cycle, is given as
\beq
\overline{ V} = \f{\omega}{2 \pi} \int_0^{2\pi/\omega} V_{sph} \, dt  \sim  \f{1}{\alpha} \f{\mu'}{\mu_0} \f{1-f}{f^2}\, . \nonumber 
\eeq
Since $f \ge 1$, the body burrows downward in the vertical direction. 

In order to maximize digging velocity for a given geometry we set, 
\beq
0=  \Dp{\overline{ V}}{  (S_{sph}/S_{cyl} ) } \propto \f{f- 2 }{f^3}  
\eeq
indicating that the maximum burrowing velocity occurs when $f =2$ or equivalently when $ {S_{sph}} = { S_{cyl} }$ regardless of the value of $\alpha$ or $\mu'$.  Thus, with $S_{cyl}= 2 \pi a_0 L$ and $S_{sph} = 4 \pi a_0^2$ the maximum velocity occurs at $a_0/L = 0.5$.  However, this optimal aspect ratio depends on our choice of $S$, which depends on the details of the burrowing system. 

\subsection{Limiting case 2: Low Reynolds number Newtonian flows}
As the Reynolds number approaches zero, forces acting on the bodies are linearly proportional to velocity ($\alpha = 1$).  Expressions for these forces can be derived analytically, in particular, for an infinitely long cylinder aligned with the flow, the drag force is given by $2 \pi \mu_f V L$ and the drag force on a sphere is given by  $6 \pi \mu_f V a_0$.  
We again consider the limit of small resistance perturbations $\mu'$ due to local changes in particle packing fraction in a viscous fluid.

The calculation is the same as the previous calculation with the exception that $S_{sph} = a_0$ and  $S_{cyl} = L/3$.  Hence the corresponding burrowing velocity is maximized at
$a_0/L = 1/3$, corresponding to a more elongated cylinder than in the previous analysis.\footnote{Note that there is a subtlety in this calculation that needs to be addressed.  The previous calculation, which should also be relevant at low Reynolds numbers, yielded an optimal aspect {\color{blue} ratio} of 1/2, not 1/3.  To rationalize this apparent discrepancy, consider the drag on a sphere in a low Reynolds number flow.  There are two common (and equivalent) ways to express the drag force: $F_D = 6 \pi \mu_f V a_0$ or $F_D = C_D \rho/2 V^2 S$ where $C_D = 24/Re$ and $S = \pi a_0^2$.  In the first case, in our formulation, we set $\mu = 6 \pi \mu_f$ and $S_{sph} = a_0$ resulting in an optimal aspect ratio of 1/3.  In the second, we set $\mu =  C_D \rho/2$ and $S_{sph} = S$ yielding an optimal aspect ratio of 1/2.  To determine which is correct, we need to consider the constitutive relationship for the resistance coefficient.  If $6 \pi \mu_f$ can be well-approximated as a linear function of $\epsilon$, the first formulation is relevant.  If, on the other hand $C_D \rho/2 = 6 \mu_f/(VR)$ (which includes the geometric parameter $R$ which also affects the void fraction) can be considered a linear function of $\epsilon$, the second estimate is more appropriate.}

{\color{blue}
\subsection{Near random close packing}
For simplicity, our model has been linearized assuming small variations in the packing fraction. 
This reduced model provides a number of general insights into how reciprocal motion of non-symmetric bodies can generate unidirectional motion in a saturated soil.
However, as the packing fraction of the substrate approaches critical transitional values (i.e. approaching random close packing), a fully nonlinear viscosity model is more appropriate. 

There are a number of effective viscosity models which correspond to the special case of $\alpha = 1$ associated with burrowing at low Reynolds numbers. 
A few such models relating the effective viscosity to the packing fraction, $\phi$, are itemized in the table below. 
\begin{center}
\begin{tabular}{|c|c|} \hline
$\mu/\mu_f$ & Reference \\ \hline \hline
$1+ 2.5 \phi$ &  Einstein \cite{Einstein1906} \\  \hline
$1+ 2.5 \phi + 7.6 \phi^2$ & Batchelor \& Green \cite{BG1972}  \\ \hline
$\left( \frac{9}{8} \right) \frac{(\phi/\phi_m)^{1/3}}{ 1- (\phi/\phi_m)^{1/3}} $ &  Frankel \& Acrivos \cite{FA1967} \\ \hline 
$\left( 1 - \frac{\phi}{\phi_m} \right)^{-\eta \phi_m}$ & Krieger \& Dougherty \cite{KD1959} \\ \hline
\end{tabular}
\end{center}

\noindent 
If we repeat the previous calculation to determine average digging velocities, this time taking into account the  full nonlinear dependence cited in \cite{KD1959} and using typical parameters ($\eta = 2.5 $, $\phi_m = 0.67$), the resultant velocity becomes
$\overline{V} = 0.1295$, which is commensurate with the  values predicted by the linearized theory.
}

\subsection{Previous numerical results}
While to the best of our knowledge this paper represents the first theoretical analysis of a simple burrower using local fluidization to propel itself, this type of digging strategy has previously been studied numerically by Shimada {\em et.~al.} \cite{Shi:09}.
In that study, the authors used an event-driven granular simulation to model a ``pushme-pullyou" consisting of two expanding and contracting disks separated by a spring.  Both halves of the body were disks, hence aspect ratio was not a parameter in their study.   
In both studies (present and previous numerics) the burrowing velocity was found to be proportional to $\omega$ at low frequencies.  In the event-driven simulations, the authors found that this relation peaks at a critical frequency and the velocity declines beyond this critical value.  
Our current theory is unable to predict these nonlinearities observed at high $\omega$ on account of the assumptions made in the constitutive relationships, namely $\mu \propto \epsilon$.  A more realistic constitutive model (which would depend on the details of soil type, preparation, etc.) is likely to exhibit behavior that is qualitatively similar to the numerical simulations at high frequencies.

\subsection{Conclusions}

In this paper, we introduce a simple theoretical model to capture key physical aspects of burrowing {\it Ensis} and other biological or engineered burrowing systems.  Even though the cylinder and sphere motion is actuated reciprocally in an over-damped environment, net unidirectional motion is achieved because of varying drag on the bodies owing to local changes in void fraction.  We find that burrowing velocities depend on the aspect ratio, $a_0/L$, and that the ``best" aspect ratio (i.e.~the one that maximizes the velocity or the efficiency) depends on the geometric details of the drag force expression.  It is interesting to note that, while we found an optimal ratio of 1/3 for viscously dominated substrates, live razor clams have an  aspect ratio closer to $a_0/L \sim 1/6$.  This discrepancy is likely to arise due to an over-simplification of the constitutive relationships describing the substrate or may be an indication that razor clams have not evolved to maximize digging speeds.  
To better describe the dynamics of {\it Ensis}, future work will focus on more precisely determining the parameters and $\epsilon$ dependence in the force relation expressed in equation (\ref{eq:F}) and investigating other cost functions.


\bibliographystyle{elsarticle-num}

\end{document}